\documentclass[aps,floats,nofootinbib,amssymb,preprint,superscriptaddress]{revtex4}
\usepackage{graphicx}

\begin{document}

\title{Charged Higgs phenomenology in the lepton-specific 
two Higgs doublet model}

\author{Heather~E.~Logan}
\email{logan@physics.carleton.ca}
\affiliation{Ottawa-Carleton Institute for Physics, Carleton University,
Ottawa K1S 5B6 Canada}

\author{Deanna MacLennan}
\affiliation{Ottawa-Carleton Institute for Physics, Carleton University,
Ottawa K1S 5B6 Canada}

\begin{abstract}
We study the ``lepton-specific'' two Higgs doublet model, in which one
doublet $\Phi_{\ell}$ gives mass to charged leptons and the other
$\Phi_q$ gives mass to both up- and down-type quarks.  We examine the
existing experimental constraints on the charged Higgs boson mass and
the parameter $\tan\beta \equiv \langle \Phi_q^0 \rangle / \langle
\Phi_{\ell}^0 \rangle$.  The most stringent constraints come from LEP-II
direct searches and lepton flavour universality in $\tau$ decays.
The former yields
$M_{H^{\pm}} \geq 92.0$~GeV; the latter yields two allowed regions,
$0.61 \tan\beta \ {\rm GeV} \leq M_{H^{\pm}} \leq 0.73 \tan\beta$ GeV or
$M_{H^{\pm}} \geq 1.4 \tan\beta$~GeV, and excludes parameter
regions beyond the LEP-II bound for $\tan\beta \gtrsim 65$.
We present the charged Higgs decay branching
fractions and discuss prospects for charged Higgs discovery at the LHC
in this model.
\end{abstract}

\thispagestyle{empty}
\maketitle


\section{Introduction}

While the Standard Model (SM) of electroweak interactions has been
rigourously tested over the past two decades, the dynamics of
electroweak symmetry breaking have yet to be probed directly.  This
leaves open the possibility of an extended Higgs sector more
complicated than the single SU(2) doublet present in the SM.

Models with two Higgs doublets (2HDMs) have been studied extensively.
In particular, the Type-II
2HDM~\cite{Lee:1973iz,Fayet:1974fj,Peccei:1977hh,Fayet:1976cr}, in
which one doublet generates the masses of up-type quarks while the
other generates the masses of down-type quarks and charged
leptons, arises naturally in supersymmetric models; its collider
phenomenology has received much attention.  The Type-I
2HDM~\cite{Georgi:1978wr,Haber:1978jt}, in which one doublet generates
the masses of all quarks and leptons while the other 
contributes only to the $W$ and $Z$ boson masses, has also been widely
considered, particularly in the context of indirect constraints.
Other patterns of couplings of two Higgs doublets to SM fermions have
been introduced~\cite{Barnett:1983mm,Barnett:1984zy,Grossman:1994jb},
but their phenomenology has not been extensively explored.

In this paper we study the ``lepton-specific'' two Higgs doublet
model\footnote{In the literature, this scenario has also been referred
to as Model IIA~\cite{Barnett:1983mm,Barnett:1984zy}, Model
I$^{\prime}$~\cite{Grossman:1994jb}, the leptonic
Higgs~\cite{Goh:2009wg}, the Type-X 2HDM~\cite{Aoki:2009ha}, and the
leptophilic 2HDM~\cite{Su:2009fz}.}, in which one doublet
$\Phi_{\ell}$ generates the masses of the charged leptons while the
second doublet $\Phi_q$ generates the masses of both up- and down-type
quarks.  This coupling structure was first introduced in
Refs.~\cite{Barnett:1983mm,Barnett:1984zy,Grossman:1994jb}, and
initial studies of the Higgs boson couplings and their detection
prospects at the CERN Large Electron Positron (LEP) collider were made
in Ref.~\cite{AkeroydLEP}.  Further studies of the couplings, decays,
and phenomenology at the CERN Large Hadron Collider (LHC) of mainly
the neutral Higgs bosons in this model have been made in
Refs.~\cite{Akeroyd:1998ui,BrooksThomas,Hmodels,Goh:2009wg,Aoki:2009ha,Su:2009fz}.
This doublet structure was also introduced in Ref.~\cite{Aoki:2008av}
(along with additional SU(2) singlet scalars) in order to avoid the
stringent constraints on the charged Higgs mass from $b \to s
\gamma$~\cite{Misiak:2006zs} that arise in the usual Type-II 2HDM.

We focus here on the charged Higgs boson $H^{\pm}$.  We study the
existing experimental constraints on the charged Higgs mass from
direct searches as well as indirect constraints on the mass and
couplings from virtual charged Higgs exchange in both tree-level and
one-loop processes.  Because of the structure of the Yukawa
Lagrangian, couplings of $H^{\pm}$ to leptons are enhanced by a factor
of $\tan\beta \equiv \langle \Phi_q^0 \rangle / \langle \Phi_{\ell}^0
\rangle$, while couplings of $H^{\pm}$ to quarks contain a
factor of $\cot\beta$.  This leads to different indirect constraints
and charged Higgs decay branching fractions than in the usual Type-I
or II 2HDMs.

This paper is organized as follows. In Sec.~\ref{sec:model} we outline
the model and present the relevant Feynman rules for the couplings of
the charged Higgs to fermions. In Sec.~\ref{sec:exptconstraints} we
present the constraints on the charged Higgs sector from direct
searches at LEP-II as well as indirect constraints from virtual
charged Higgs exchange.  The most stringent indirect constraint comes
from $\mu$--$e$ universality in $\tau$ decays.  We also review the
charged Higgs effects in muon and $\tau$ decay distributions, $B^+$
and $D_s^+$ leptonic decays, $b \to c \tau\nu$, $B_{(s)} \to
\ell^+\ell^-$, and $b\to s\gamma$.  In Sec.~\ref{sec:decaybrs} we plot
the decay branching fractions of $H^{\pm}$ as a function of the
charged Higgs mass for various values of $\tan\beta$ and compare them
to those in the usual Type-II 2HDM. We finish in
Sec.~\ref{sec:discussion} with a discussion of charged Higgs search
prospects at the LHC and a summary of our conclusions.


\section{The Model}
\label{sec:model}

We begin with two complex SU(2)-doublet fields $\Phi_q$ and $\Phi_{\ell}$,
with
\begin{equation}
   \Phi_i = \left( \begin{array}{c}
\phi_i^+ \\
\frac{1}{\sqrt{2}} \left( \phi_i^{0,r} + v_i + i \phi_i^{0,i} \right) 
\end{array} \right), \qquad\qquad 
i = q, \ell.
\end{equation}
The structure of the Yukawa Lagrangian that characterizes this model is 
enforced by imposing a discrete symmetry under which $\Phi_{\ell}$ and
the right-handed leptons transform as
\begin{equation}
  \Phi_{\ell} \rightarrow -\Phi_{\ell}, \qquad \qquad 
  e_{Ri} \rightarrow -e_{Ri},
\end{equation}
while all other fields remain unchanged.  The resulting Yukawa Lagrangian
is
\begin{equation}
  \mathcal{L}_{\rm Yuk} = -\sum_{i,j=1}^3 \left[ 
   y_{ij}^{u} \overline{u}_{Ri} \widetilde{\Phi}_q^{\dagger} Q_{Lj}
 + y_{ij}^{d} \overline{d}_{Ri} \Phi_{q}^{\dagger} Q_{Lj}
 + y_{ij}^{\ell} \overline{e}_{Ri} \Phi_{\ell}^{\dagger} L_{Lj} \right]
 + {\rm h.c.},
\label{eq:Lyuk}
\end{equation}
where $i,j$ are generation indices, $y^{u,d,\ell}_{ij}$ are the Yukawa 
coupling matrices, the left-handed quark and lepton doublets are
\begin{equation}
  L_{Li} = \left( \begin{array}{c}
    \nu_{Li} \\
    e_{Li}
    \end{array} \right), \qquad 
  Q_{Li} = \left( \begin{array}{c}
  u_{Li} \\
  d_{Li} 
  \end{array} \right),
\end{equation}
and the conjugate Higgs doublet is given by
\begin{equation}
  \widetilde{\Phi}_q \equiv i \sigma_2 \Phi_q^*
  = \left( \begin{array}{c}
  \frac{1}{\sqrt{2}} \left( \phi_q^{0,r} + v_q - i \phi_q^{0,i} \right) \\
  -\phi_q^- 
  \end{array} \right).
\end{equation}

The charged states $\phi^+_q$ and $\phi^+_{\ell}$ mix to form the charged
Goldstone boson and a single physical charged Higgs state,
\begin{equation}
  H^+ = - \phi^+_{\ell} \sin\beta + \phi_q^+ \cos\beta,
\end{equation}
where we define $\tan\beta = v_q/v_{\ell}$.  We also have 
$\sqrt{v_q^2 + v_{\ell}^2} = v_{\rm SM} = 2 M_W/g \simeq 246$ GeV,
where $M_W$ is the $W$ boson mass and $g$ is the SU(2) gauge coupling.

The Feynman rules for charged Higgs boson couplings to fermions are 
given as follows, with all particles incoming:\footnote{For comparison, 
the corresponding couplings in the Type-I 2HDM are~\cite{HHG}
\begin{eqnarray}
  H^+ \overline{u}_i d_j & \ : \ & \frac{ig}{\sqrt{2} M_W} V_{ij} 
  \cot\beta \left( m_{u_i} P_L - m_{d_j} P_R \right), \nonumber \\
  H^+ \overline{\nu}_{e_k} e_k & \ : \ & -\frac{ig}{\sqrt{2} M_W} 
  \cot\beta \, m_{e_k} P_R,
\end{eqnarray}
with $\tan\beta = v_2/v_1$ where $v_2$ is the vacuum expectation value
of the Higgs field that couples to fermions; the other doublet is decoupled
from fermions.
In the Type-II 2HDM the couplings are~\cite{HHG}
\begin{eqnarray}
  H^+ \overline{u}_i d_j & \ : \ & \frac{ig}{\sqrt{2} M_W} V_{ij} 
  \left( \cot\beta \, m_{u_i} P_L + \tan\beta \, m_{d_j} P_R \right), 
\nonumber \\
  H^+ \overline{\nu}_{e_k} e_k & \ : \ & \frac{ig}{\sqrt{2} M_W} 
  \tan\beta \, m_{e_k} P_R,
\end{eqnarray}
again with $\tan\beta = v_2/v_1$; this time $v_1$ ($v_2$) is the vacuum
expectation value of the doublet that couples to down-type quarks and charged 
leptons (up-type quarks).}
\begin{eqnarray}
  H^+ \overline{u}_i d_j & \ : \ & \frac{ig}{\sqrt{2} M_W} V_{ij} 
  \cot\beta \left( m_{u_i} P_L - m_{d_j} P_R \right), \nonumber \\
  H^+ \overline{\nu}_{e_k} e_k & \ : \ & \frac{ig}{\sqrt{2} M_W} 
  \tan\beta \, m_{e_k} P_R.
  \label{eq:Feynmanrules}
\end{eqnarray}
Here $V_{ij}$ is the relevant CKM matrix element and $P_{L,R} \equiv
(1 \mp \gamma^5)/2$ are the left- and right-handed projection operators.

Note that the $H^+ \bar \nu \ell$ couplings are enhanced at large
$\tan\beta$ while the $H^+ \bar u d$ couplings are suppressed.  This
enhancement of the lepton couplings is due to the $m_{\ell}/v_{\ell}$
dependence of the lepton Yukawa couplings,
\begin{equation}
   y_{\ell} = \frac{\sqrt{2} m_{\ell}}{v_{\ell}}
   = \frac{\sqrt{2} m_{\ell}}{v_{\rm SM} \cos\beta}.
\end{equation}
The maximum value of $\tan\beta$ is limited by the requirement that
the $\tau$ Yukawa coupling remain perturbative,
\begin{equation}
  y_{\tau} = \frac{\sqrt{2} m_{\tau}}{v_{\rm SM} \cos\beta} \lesssim 4 \pi.
\end{equation}
This leads to an upper bound on $\tan\beta$ of
\begin{equation}
  \tan\beta \lesssim 1200.
\end{equation}
In our numerical results we will consider values of $\tan\beta$ up to
100 or 200, corresponding to $y_{\tau}$ values of about 1 or 2, respectively.


\section{Experimental constraints}
\label{sec:exptconstraints}

\subsection{LEP-II direct search}

The four LEP collaborations have presented combined
limits~\cite{LEPchargedHiggs} for $e^+e^- \to H^+H^-$ with $H^+ \to
\tau \nu$ or $c \bar s$, assuming that the branching ratios of these
two decays add to 1.  The 95\% confidence level (CL) limits range from
$M_{H^{\pm}} \geq 78.6$~GeV to 89.6~GeV; the strongest limit is reached
for BR($H^+ \to \tau \nu) = 1$.  Separately, the OPAL collaboration
presented a charged Higgs search in the $\tau\nu\tau\nu$ channel alone
assuming BR($H^+ \to \tau \nu) = 1$, which excludes
$M_{H^{\pm}}$ values below 92.0~GeV at 95\% CL~\cite{OPALchargedHiggs}.

In this paper we are interested in $\tan\beta$ values greater than a
few.  In this case, as we will show in Sec.~\ref{sec:decaybrs}, the
branching ratio of $H^+ \to \tau \nu$ is very close to 1 for charged
Higgs masses in the region of the LEP-II limit.  We therefore take the
more stringent OPAL limit~\cite{OPALchargedHiggs},
\begin{equation}
  M_{H^{\pm}} \geq 92.0 \ {\rm GeV}.
  \label{eq:lep2bound}
\end{equation}

\subsection{Lepton universality in $\tau$ decays}

The decays $\tau \to \mu \bar \nu_{\mu} \nu_{\tau}$, $\tau \to e \bar
\nu_e \nu_{\tau}$, and $\mu \to e \bar \nu_e \nu_{\mu}$ proceed at
tree level in the SM through virtual $W$ exchange.  In models with two
Higgs doublets they also receive a contribution from tree-level
charged Higgs exchange.  The tree-level partial width for these decays
in the lepton-specific 2HDM is identical to that in the Type-II 
2HDM~\cite{KrawczykPokorski,HollikSack},
\begin{equation}
  \Gamma(L \to \ell \bar \nu_{\ell} \nu_L) = \frac{G_F^2 m_L^5}{192 \pi^3}
  \left[ \left( 1 
  + \frac{1}{4}m_{\ell}^2 m_L^2 \frac{\tan^4\beta}{M_{H^{\pm}}^4} \right)
  f(m_{\ell}^2/m_L^2)
  - 2 m_{\ell}^2 \frac{\tan^2\beta}{M_{H^{\pm}}^2} 
  g(m_{\ell}^2/m_L^2) \right],
  \label{eq:Ltolnunu}
\end{equation}
where here $L$ denotes the initial lepton, $\ell$ denotes the
final-state charged lepton, and the phase space factors $f$ and $g$ are
given by~\cite{HollikSack}
\begin{equation}
  f(x) = 1 - 8 x + 8 x^3 - x^4 - 12 x^2 \ln x, \qquad 
  g(x) = 1 + 9 x - 9 x^2 - x^3 + 6 x (1+x) \ln x.
\end{equation}
The two terms in the parentheses in Eq.~\ref{eq:Ltolnunu} come from
the square of the usual SM $W^{\pm}$ exchange diagram and the square of the
charged Higgs exchange diagram, respectively.  The remaining term is
the (destructive) interference between the $W^{\pm}$ diagram and the charged
Higgs diagram, which requires a helicity flip of the final state
lepton $\ell$ yielding an extra suppression factor $m_{\ell}/m_L$ and
a different phase space factor.  Because of the lepton mass
dependence, the effect of the charged Higgs exchange will be largest
in $\tau \to \mu \bar \nu \nu$.

Additional 2HDM corrections to charged lepton decay arise from
one-loop diagrams involving charged and neutral Higgs bosons
contributing to the $L \nu_L W$ and $\ell \nu_{\ell} W$
vertices~\cite{Krawczyk:2004na}.  Particularly significant are the
corrections to the $\tau \nu_{\tau} W$ vertex, because they involve
two powers of the $\tau$ Yukawa coupling and are not suppressed by the
charged Higgs coupling to muons or electrons.  These $\tau$ vertex
corrections are the same for the $\tau \to \mu \bar \nu \nu$ and $\tau
\to e \bar \nu \nu$ channels.  They also depend on the neutral Higgs
masses and mixing angle as well as $M_{H^{\pm}}$ and $\tan\beta$.  In
the present paper we focus on the charged Higgs sector alone; we
will therefore consider an observable in which the one-loop
corrections to the $\tau \nu W$ vertex cancel.

The SM $W^+ \ell \bar \nu$ couplings are generation-universal and
$\tau$ and muon decay suffer no helicity suppression.  The $H^+ \ell
\bar \nu$ couplings, on the other hand, depend on the mass of the
charged lepton involved.  Therefore, tests of flavour universality in
the couplings that mediate $\tau$ and muon decays are sensitive to
charged Higgs contributions.
The $\tau$ decay rates can be written in terms of the muon lifetime
$\tau_{\mu}$ in the standard form (see, e.g., Ref.~\cite{Roney}),
\begin{eqnarray}
  \tau_{\tau} &=& \frac{g_{\mu}^{2}}{g_{\tau}^{2}} 
  \tau_{\mu} \frac{m_{\mu}^5}{m_{\tau}^5} 
  {\rm BR}(\tau \rightarrow e \overline{\nu}_e \nu_{\tau})
  \frac{f(m_e^2/m_{\mu}^2) r_{RC}^{\mu}}{f(m_e^2/m_{\tau}^2) r_{RC}^{\tau}},
  \nonumber \\
  \tau_{\tau} &=& \frac{g_e^2}{g_{\tau}^2} 
  \tau_{\mu} \frac{m_{\mu}^5}{m_{\tau}^5} 
  {\rm BR}(\tau \rightarrow \mu \overline{\nu}_{\mu} \nu_{\tau})
  \frac{f(m_e^2/m_{\tau}^2) r_{RC}^{\mu}}
   {f(m_{\mu}^2/m_{\tau}^2) r_{RC}^{\tau}},
\end{eqnarray}
where $r_{RC}^i$ are the QED radiative corrections to the SM
decays.  Here any deviations from flavour universality are
parameterized by effective charged current couplings $g_e$, $g_{\mu},$
and $g_{\tau}$, which are equal to 1 in the SM.  Ratios of these
parameters are extracted from measurements of the $\tau$ lifetime and
the $\tau$ branching ratios to $e \bar \nu \nu$ and $\mu \bar \nu
\nu$.  The current world-average experimental values are~\cite{Roney}
\begin{equation}
  \frac{g_{\mu}}{g_e} = 0.9999 \pm 0.0020, \qquad \qquad
  \frac{g_{\mu}}{g_{\tau}} = 0.9982 \pm 0.0021.
\end{equation}

\begin{figure}
\resizebox{0.8\textwidth}{!}{\includegraphics{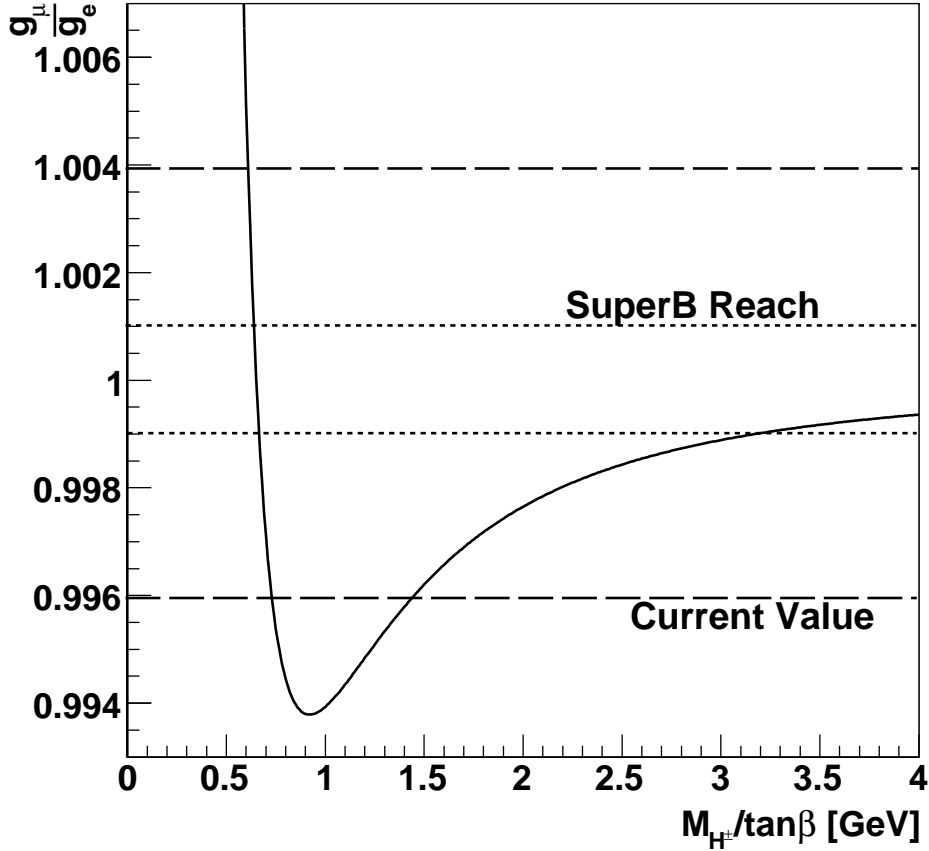}}
\caption{\label{fig:gmuge} Prediction for $g_{\mu}/g_e$ in the
lepton-specific 2HDM as a function of $M_{H^{\pm}}/\tan\beta$ (solid line).
Horizontal dashed lines indicate the current $2\sigma$ allowed range
from lepton universality in $\tau$ decays (outer lines) and the future
anticipated reach of SuperB (inner lines). }
\end{figure}

The observable $g_{\mu}/g_e$ comes from the ratio of the $\tau$ 
leptonic branching fractions.
In the lepton-specific 2HDM we have at tree level,
\begin{equation}
  \frac{g_{\mu}^2}{g_e^2} = 
  \frac{1 + m_{\mu}^2 m_{\tau}^2 \tan^4\beta / 4 M_{H^{\pm}}^4
         - (2 m_{\mu}^2 \tan^2\beta / M_{H^{\pm}}^2) 
    g(m_{\mu}^2/m_{\tau}^2)/f(m_{\mu}^2/m_{\tau}^2)}
       {1 + m_e^2 m_{\tau}^2 \tan^4\beta / 4 M_{H^{\pm}}^4
	 - (2 m_e^2 \tan^2\beta / M_{H^{\pm}}^2) 
	 g(m_e^2/m_{\tau}^2)/f(m_e^2/m_{\tau}^2)},
\end{equation}
and one-loop 2HDM corrections to the $\tau \nu W$ vertex cancel in the
ratio.\footnote{We note that the tree-level expression for the other
observable,
\begin{equation}
  \frac{g_{\mu}^2}{g_{\tau}^2} =
  \frac{1 + m_e^2 m_{\mu}^2 \tan^4\beta / 4 M_{H^{\pm}}^4
    - (2 m_e^2 \tan^2\beta / M_{H^{\pm}}^2)
    g(m_e^2/m_{\mu}^2)/f(m_e^2/m_{\mu}^2)}
       {1 + m_e^2 m_{\tau}^2 \tan^4\beta / 4 M_{H^{\pm}}^4
	 - (2 m_e^2 \tan^2\beta / M_{H^{\pm}}^2)
	 g(m_e^2/m_{\tau}^2)/f(m_e^2/m_{\tau}^2)},
\end{equation}
is very close to its SM value due to the $m_e$ factors in the charged 
Higgs exchange terms.  This observable, however, is sensitive to the 
one-loop corrections discussed in Ref.~\cite{Krawczyk:2004na} and can
be used to constrain the neutral Higgs sector of the 2HDM.  Such an 
analysis is beyond the scope of this paper.}
The square root of this ratio is plotted in Fig.~\ref{fig:gmuge}
as a function of $M_{H^{\pm}}/\tan\beta$, along
with the current $2\sigma$ experimental limits from Ref.~\cite{Roney}.
Inserting the experimental results yields two allowed regions at 95\% CL:
\begin{equation}
  0.61 \tan\beta \ {\rm GeV} \leq M_{H^{\pm}} \leq 0.73 \tan\beta \ {\rm GeV}
  \qquad {\rm or} \qquad
  M_{H^{\pm}} \geq 1.4 \tan\beta \ {\rm GeV}.
\end{equation}
This constraint begins to exclude parameter regions beyond the LEP-II
bound when $\tan\beta \gtrsim 65$.

Measurements of $\tau$ branching fractions from the proposed SuperB
high-luminosity flavour factory~\cite{SuperB} are expected to improve
the precision on $g_{\mu}/g_e$ to better than 0.05\%~\cite{Roney}.  In
the absence of a deviation from the SM prediction, this would give an
even tighter constraint on the charged Higgs mass,
\begin{equation}
  0.64 \tan\beta \ {\rm GeV} \leq M_{H^{\pm}} \leq 0.67 \tan\beta \ {\rm GeV}
  \qquad {\rm or} \qquad
  M_{H^{\pm}} \geq 3.2 \tan\beta \ {\rm GeV} \qquad \qquad {\rm (SuperB)}.
\end{equation}
Such a constraint would exclude parameter regions beyond the LEP-II bound
when $\tan\beta \gtrsim 30$.

\begin{figure}
\resizebox{0.8\textwidth}{!}{\includegraphics{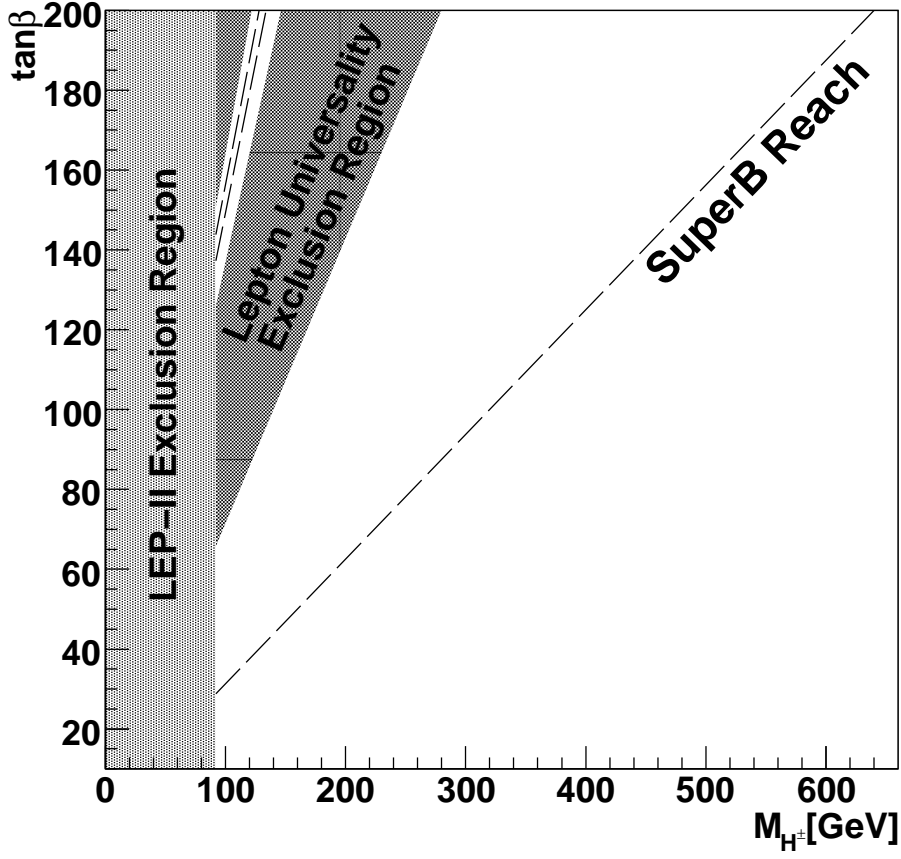}}
\caption{\label{fig:limits} Constraints on $M_{H^{\pm}}$ and
$\tan\beta$ at 95\% CL from LEP-II direct searches and lepton flavour
universality in $\tau$ decays.  The dashed lines show the anticipated
reach of the SuperB experiment.  Note the allowed sliver of parameter
space at lower $M_{H^{\pm}}/\tan\beta$.}
\end{figure}

The constraints on $M_{H^{\pm}}$ and $\tan\beta$ due to LEP-II direct
searches and flavour universality in $\tau$ decays are summarized in
Fig.~\ref{fig:limits}.

\subsection{Other low-energy processes}

\subsubsection{Michel parameters in muon and $\tau$ decay}

In the SM, muon and $\tau$ decays proceed through the left-handed vector
couplings of the $W$ boson.  The $H^+$ exchange contribution in the
lepton-specific 2HDM involves scalar couplings to right-handed charged
leptons (Eq.~\ref{eq:Feynmanrules}).  This different coupling
structure can affect the energy and angular distribution of the
daughter charged lepton in decays of polarized muons or $\tau$s.
These distributions are parameterized in terms of the Michel 
parameters~\cite{Michel} $\rho$, $\xi$, $\delta$, and $\eta$, which
are defined in terms of 
the energy and angular distribution of the daughter charged lepton
$\ell^{\pm}$ in the rest
frame of the parent ($L^{\pm}$)~\cite{PDGmuonreview}:
\begin{eqnarray}
  \frac{d^2\Gamma}{dx \, d\cos\theta} &\propto& x^2 \left\{
  3(1-x) + \frac{2 \rho}{3}(4x-3) + 3 \eta x_0 (1-x)/x \right.
\nonumber \\
  && \left. \pm P \xi \cos\theta \left[
  1 - x + \frac{2\delta}{3} (4x - 3) \right] \right\}.
\end{eqnarray}
Here $\theta$ is the angle between the $\ell^{\pm}$ momentum and the
parent lepton's spin, $x = 2E_{\ell}/m_L$, $x_0 = 2 m_{\ell}/m_L$,
$P$ is the degree of polarization of the parent lepton, and we have
neglected neutrino masses and terms higher order in $m_{\ell}/m_L$.
The SM values for the Michel parameters are $\rho = 3/4$, $\xi = 1$,
$\eta = 0$, and $\delta = 3/4$.

The most general expression for the Michel parameters is given
by~\cite{FetscherGerber}
\begin{eqnarray}
  \rho &=& \frac{3}{4} - \frac{3}{4} \left[|g_{RL}^V|^2 + |g_{LR}^V|^2 
  + 2|g_{RL}^T|^2 + 2|g_{LR}^T|^2 
  + Re\left(g_{RL}^S g_{RL}^{T*} + g_{LR}^S g_{LR}^{T*} \right) \right],
  \nonumber \\
  \eta &=& \frac{1}{2} Re\left[g_{RR}^V g_{LL}^{S*} + g_{LL}^V g_{RR}^{S*}
  + g_{RL}^V \left(g_{LR}^{S*} + 6 g_{LR}^{T*}\right)
  + g_{LR}^V \left(g_{RL}^{S*} + 6g_{RL}^{T*}\right) \right], \nonumber \\
  \xi &=& 1 - \frac{1}{2} |g_{LR}^S|^2 - \frac{1}{2}|g_{RR}^S|^2
  - 4 |g_{RL}^V|^2 + 2 |g_{LR}^V|^2 - 2 |g_{RR}^V|^2 \nonumber \\
  & & + 2 |g_{LR}^T|^2 - 8 |g_{RL}^T|^2 
  + 4 Re\left(g_{LR}^S g_{LR}^{T*} - g_{RL}^S g_{RL}^{T*}\right), \nonumber \\
  \xi\delta &=& \frac{3}{4} - \frac{3}{8} |g_{RR}^S|^2 
  - \frac{3}{8} |g_{LR}^S|^2 - \frac{3}{2} |g_{RR}^V|^2 
  - \frac{3}{4} |g_{RL}^V|^2 - \frac{3}{4} |g_{LR}^V|^2 \nonumber \\
  & & -\frac{3}{2} |g_{RL}^T|^2 - 3 |g_{LR}^T|^2 
  + \frac{3}{4} Re\left(g_{LR}^S g_{LR}^{T*} - g_{RL}^S g_{RL}^{T*}\right),
\end{eqnarray}
where the couplings are defined in terms of the most general matrix 
element for the charged lepton decay $L^- \to \ell^- \bar \nu_{\ell} \nu_L$
according to~\cite{Scheck} 
\begin{equation}
  \mathcal{M} = 4 \frac{G_{F}}{\sqrt{2}} \sum_{\gamma=S,V,T} 
  \sum_{\alpha,\beta=R,L} g_{\alpha\beta}^{\gamma}
  \left\langle \overline{\ell}_{\alpha} \left| \Gamma^{\gamma} \right|
  \nu_{\ell} \right\rangle
  \left\langle \overline{\nu}_L \left| \Gamma_{\gamma} \right|
  L_{\beta} \right\rangle.
\end{equation}
Here $\gamma = S$, $V$, or $T$ denotes scalar ($\Gamma^S = 1$), vector
($\Gamma^V = \gamma^{\mu}$), or tensor ($\Gamma^T =
\sigma^{\mu\nu}/\sqrt{2} = i[\gamma^{\mu},\gamma^{\nu}]/2\sqrt{2}$) 
interactions, respectively, and the
chiralities of $\ell$ and $L$ are specified by $\alpha$ and $\beta$,
respectively.

We consider the decay $L \to \ell \bar \nu \nu$ where $L$ ($\ell$)
is replaced by $\mu$ ($e$) for muon decay and by $\tau$ ($\mu$ or $e$)
for $\tau$ decay.  In the lepton-specific 2HDM, we have $g_{LL}^V =
-1/4$ representing SM $W$ boson exchange and $g_{RR}^S = m_L m_{\ell}
\tan^2\beta / 4 M_{H^{\pm}}^2$ representing charged Higgs exchange.
All other couplings $g^{\gamma}_{\alpha\beta}$ are zero.
The Michel parameters become
\begin{eqnarray}
  \rho &=& \frac{3}{4}, \nonumber \\
  \eta &=& -\frac{m_L m_{\ell}}{32}
       \frac{\tan^2\beta}{M_{H^{\pm}}^2}, \nonumber \\
  \xi &=& 1 - \frac{m_L^2 m_{\ell}^2}{32}
       \frac{\tan^4\beta}{M_{H^{\pm}}^4}, \nonumber \\
  \xi \delta &=& \frac{3}{4} \left[1 - \frac{m_L^2 m_{\ell}^2}{32}
       \frac{\tan^4\beta}{M_{H^{\pm}}^4} \right]
       = \frac{3}{4} \xi.
\end{eqnarray}
The parameters $\rho$ and $\delta$ are equal to their SM values and
provide no constraints.  

We summarize the constraints on $M_{H^{\pm}}$ and $\tan\beta$ from the
Michel parameters in muon and $\tau$ decay in Table~\ref{tab:michel}.
The strongest constraint comes from $\eta$ and $\xi$ in $\tau \to \mu
\bar\nu \nu$, which coincidentally yield the same limit at 95\% CL:
\begin{equation}
  M_{H^{\pm}} \geq 0.34 \tan\beta \ {\rm GeV}.
\end{equation}
This constraint is not competitive with that from lepton flavour
universality in $\tau$ decays.  We note that an improvement in the
$2\sigma$ lower bound on $\eta$ ($\xi$) in $\tau \to \mu \bar\nu \nu$
decay to $-0.010$ ($0.996$) would be required to raise this limit to
$M_{H^{\pm}} \geq 0.73 \tan\beta$~GeV and eliminate the allowed sliver
of parameter space at lower $M_{H^{\pm}}/\tan\beta$ values from
current data on lepton universality in $\tau$ decays (see
Fig.~\ref{fig:limits}).

\begin{table}
\begin{tabular}{lll}
\hline \hline
Process & Observable & Constraint \\
\hline
$\mu \to e \bar\nu \nu$ \ \ & $\eta = 0.001 \pm 0.024$ \ \ 
  & $M_{H^{\pm}} \geq 0.006 \tan\beta$ GeV \\
\hline
$\tau \to \mu \bar\nu \nu$ & $\eta = 0.094 \pm 0.073$
  & $M_{H^{\pm}} \geq 0.34 \tan\beta$ GeV \\
 & $\xi = 1.030 \pm 0.059$
  & $M_{H^{\pm}} \geq 0.34 \tan\beta$ GeV \\
\hline
$\tau \to e \bar\nu \nu$ & $\xi = 0.994 \pm 0.040$
  & $M_{H^{\pm}} \geq 0.023 \tan\beta$ GeV \\
\hline \hline
\end{tabular}
\caption{\label{tab:michel} Current world-average values of the Michel
parameters in muon and tau decay from Ref.~\cite{PDG} and the
resulting 95\% CL constraints on $M_{H^{\pm}}$ and $\tan\beta$.  (No
separate measurement of $\xi$ in muon decay or of $\eta$ in $\tau \to
e \bar\nu \nu$ is quoted in Ref.~\cite{PDG}.)}
\end{table}


\subsubsection{$B^{+} \rightarrow \tau^+ \nu_{\tau}$}

In the Standard Model, the partial width for the decay $B^+ \to \tau^+
\nu_{\tau}$ mediated by tree-level $W^+$ exchange is given by
\begin{equation}
  \Gamma_{\rm SM} (B^+ \to \tau^+ \nu_{\tau}) = 
  \frac{G_F^2}{8\pi} f_{B^+}^2 m_{B^+} m_{\tau}^2 |V_{ub}|^2
  \left[1 - \frac{m_{\tau}^2}{m_{B^+}^2}\right]^2,
\end{equation}
where $m_{B^+}$ is the $B^+$ meson mass, $V_{ub}$ is the relevant
CKM matrix element, and $f_{B^+}$ is the $B^+$ meson decay constant
defined according to
\begin{equation}
  i f_{B^+} p_{\mu} = \left\langle 0 \left| \overline{b} \gamma_{\mu} 
  \gamma_5 u \right| B^+(p) \right\rangle.
\end{equation}
The partial width is proportional to $m_{\tau}^2$ because of helicity
suppression and the term in the square brackets arises from the phase
space. 

In the lepton-specific 2HDM this decay receives an additional 
contribution from tree-level charged Higgs exchange; the total width
becomes
\begin{equation}
  \Gamma (B^+ \to \tau^+ \nu_{\tau}) = 
  \left[1 - \frac{m_{B^+}^2}{M_{H^{\pm}}^2}\right]^2
  \Gamma_{\rm SM} (B^+ \to \tau^+ \nu_{\tau}).
\end{equation}
Here the helicity suppression of the SM decay ensures that the charged
Higgs contribution contains no additional factors of $m_{\tau}$.  Note
that the contributions from $W^+$ and $H^+$ exchange interfere
destructively.
Note also that this result differs from that in the Type-II
2HDM~\cite{Hou:1992sy},
\begin{equation}
  \Gamma (B^+ \to \tau^+ \nu_{\tau}) = 
  \left[1 - \tan^2\beta \frac{m_{B^+}^2}{M_{H^{\pm}}^2}\right]^2
  \Gamma_{\rm SM} (B^+ \to \tau^+ \nu_{\tau})
  \qquad \qquad ({\rm Type \ II \ 2HDM}),
\end{equation}
which has been used to constrain $M_{H^{\pm}}/\tan\beta$ in that model
(for recent results see Ref.~\cite{akeroyd}).
In the lepton-specific 2HDM there is no $\tan^2\beta$ enhancement of
the charged Higgs contribution because while the charged Higgs
coupling to leptons is proportional to $\tan\beta$, its coupling to
quarks is proportional to $\cot\beta$ (Eq.~\ref{eq:Feynmanrules}).
Without the $\tan^2\beta$ enhancement, the contribution due to charged
Higgs exchange yields only a weak bound on $M_{H^{\pm}}$.

The allowed charged Higgs mass values can be extracted according
to\footnote{The only place that other nonstandard effects could creep
in to this expression is through $|V_{ub}|$, which is extracted from a SM
fit to many $b$ observables.  However, we expect nonstandard effects
from the lepton-specific 2HDM to be negligible in this fit because the
quark Yukawa couplings are all proportional to $\cot\beta$ and thus 
suppressed for $\tan\beta > 1$.}
\begin{equation}
  \left[ 1 - \frac{m_{B^+}^2}{M_{H^{\pm}}^2} \right]^2 
  = \frac{8 \pi \ {\rm BR}(B^+ \to \tau^+ \nu)}{\tau_{B^+} f_{B^+}^2
    G_F^2 m_{B^+} m_{\tau}^2 |V_{ub}|^2 (1 - m_{\tau}^2/m_{B^+}^2)^2},
  \label{eq:B+extract}
\end{equation}
where $\tau_{B^+}$ is the $B^+$ lifetime.  All quantities in
Eq.~\ref{eq:B+extract} have been measured experimentally except for
$f_{B^+}$, which can be taken from recent unquenched lattice
QCD results~\cite{Gray:2005ad}:
\begin{equation}
  f_{B^+} = f_B = 0.216 \pm 0.022 \ {\rm GeV}.
\end{equation}
The current world average experimental value for the branching ratio
$B^+ \rightarrow \tau^+ \nu_{\tau}$ from the BELLE and BABAR
collaborations is~\cite{hfag},
\begin{equation}
  {\rm BR}(B^+ \rightarrow \tau^+ \nu_{\tau}) 
  = \left(1.41_{-0.42}^{+0.43}\right) \times 10^{-4}.
\end{equation}
The only other quantity in Eq.~\ref{eq:B+extract} with a
non-negligible uncertainty is $|V_{ub}|$, for which we take the global
SM fit value~\cite{PDG},
\begin{equation}
  |V_{ub}| = (3.59 \pm 0.16) \times 10^{-3}.
\end{equation}
Combining all uncertainties in quadrature we obtain\footnote{For the 
other parameters we use
$\tau_{B^+} = (1.638 \pm 0.011) \times 10^{-12}$ s, 
$G_F = 1.16637(1) \times 10^{-5}$ GeV$^{-2}$,
$m_{B^+} = 5.27915(31)$ GeV, and
$m_{\tau} = 1.77684(17)$ GeV~\cite{PDG}.}
\begin{equation}
  \left[1-\frac{m_{B^+}^2}{M_{H^{\pm}}^2}\right]^2 = 1.33 \pm 0.50,
\end{equation}
which yields two allowed ranges for the charged Higgs mass at 95\% CL:
\begin{equation}
  0.63 \, m_{B^+} \leq M_{H^{\pm}} \leq 0.80 \, m_{B^+} \qquad 
  {\rm or} \qquad
  M_{H^{\pm}} \geq 1.5 \, m_{B^+} = 8.1 \ {\rm GeV}.
\end{equation}
The lower mass window is excluded by direct searches; the remaining
limit is well below the LEP-II direct search bound
(Eq.~\ref{eq:lep2bound}) and thus provides no new information.

\subsubsection{$D_s^+ \rightarrow \tau^+ \nu$}

The leptonic decay $D_s^+ \to \tau^+ \nu$ is completely analogous to
$B^+ \to \tau^+ \nu$ with the $B^+$ meson ($\bar b u$) replaced by the
$D_s^+$ meson ($\bar s c$).  The current experimental value of the
$D_s^+ \to \tau^+ \nu$ branching fraction is~\cite{PDG}
\begin{equation}
  {\rm BR}(D_s^+ \to \tau^+ \nu) = (6.6 \pm 0.6)\%
\end{equation}
and the current unquenched lattice QCD result for $f_{D_s}$
is~\cite{Follana:2007uv}
\begin{equation}
  f_{D_s} = 0.241 \pm 0.003 \ {\rm GeV}.
\end{equation}
Combining all uncertainties in quadrature as in the previous section
we obtain\footnote{For the remaining parameters we use $|V_{cs}| =
0.97334 (23)$ (global SM fit value), $m_{D_s^+} = 1.96849 (34)$
GeV, and $\tau_{D_s^+} = (500 \pm 7) \times 10^{-15}$ s~\cite{PDG}.}
\begin{equation}
  \left[1-\frac{m_{D_s^+}^2}{M_{H^{\pm}}^2}\right]^2 = 1.37 \pm 0.13.
\end{equation}
In particular, there is about a 40\% (or $3\sigma$) discrepancy
between the SM prediction and the experimental
measurement\footnote{Ref.~\cite{Dobrescu:2008er} finds a $3.8\sigma$
discrepancy after including $D_s^+ \to \mu^+ \nu$ data.}; moreover,
the branching fraction of $D_s^+ \to \tau^+ \nu$ is larger than the SM
prediction.  Because the $W^+$ and $H^+$ exchange diagrams interfere
destructively in the lepton-specific 2HDM, an explanation of the
discrepancy in this context would require the decay amplitude to be
dominated by the charged Higgs contribution, leading at 95\% CL to
\begin{equation}
  M_{H^{\pm}} = (0.68 \pm 0.01) \, m_{D_s^+} = 1.34 \pm 0.02 \ {\rm GeV}.
\end{equation}
This is clearly excluded by direct searches; moreover, such a light
charged Higgs in this model would yield sizeable effects in $B^+ \to
\tau^+ \nu$.  The discrepancy thus cannot be explained in the context
of the lepton-specific 2HDM.

We note here that, in the absence of a deviation from the SM
prediction, the current $D_s^+ \to \tau^+ \nu$ branching fraction
measurement precision would yield the allowed regions 
$0.69 \, m_{D_s^+} \leq M_{H^{\pm}} \leq 0.73 \, m_{D_s^+}$ or
$M_{H^{\pm}} \geq 3.2 \, m_{D_s^+} = 6.2$ GeV at 95\% CL.  This
measurement would thus provide a weaker constraint even than $B^+ \to
\tau^+ \nu$ at the current level of experimental uncertainty.


\subsubsection{Other $B$ decays}

Other $b$ quark decay processes have been used to constrain 2HDMs.
In the lepton-specific 2HDM, however, they do not provide useful 
constraints at moderate to large $\tan\beta$.
We discuss them briefly here.

The decay $b \to c \tau \nu$ receives a contribution from tree-level
$H^+$ exchange~\cite{Hou:1992sy,Grossman:1994ax,Grossman:1995yp}.
However, as in the case of $B^+ \to \tau^+ \nu$, the $\tan\beta$
enhancement in the $\tau$ Yukawa coupling is cancelled by the
$\cot\beta$ dependence of the quark Yukawa couplings, leading to very
small charged Higgs effects, equivalent to those in the Type-II 2HDM
with $\tan\beta = 1$.

The decay $B^0_{(s)} \to \ell^+ \ell^-$ receives corrections in the
Type-II 2HDM enhanced by $\tan^2\beta$~\cite{Bsll-2HDM}.  In the
lepton-specific 2HDM, however, there is no $\tan\beta$ enhancement,
again because the $\tan\beta$ from the lepton Yukawa coupling is
cancelled by $\cot\beta$ factors from the quark Yukawa couplings.
The constraints from this process are thus very weak.

Finally, the charged Higgs contributions to $b \to s \gamma$ involve
couplings of the charged Higgs to quarks at both vertices, yielding
two factors of $\cot\beta$ from the quark Yukawa couplings in the
amplitude.  The prediction for this process in the lepton-specific
2HDM is in fact identical to that in the Type-I
2HDM~\cite{Hewett:1992is}.  It can be used to constrain the parameter
space at small $\tan\beta$, yielding $\tan\beta \gtrsim 4$ (2) for
$M_{H^{\pm}} = 100$~GeV (500 GeV)~\cite{Su:2009fz}, but provides no constraints
at large $\tan\beta$.\footnote{It is for this reason that the
lepton-specific 2HDM was used in the model of
Ref.~\cite{Aoki:2008av}.}


\subsection{Tevatron constraints}

The Tevatron experiments have searched for charged Higgs production in
top quark decays and set upper limits on the branching ratio for $t
\to H^+ b$ with either $H^+ \to c \bar s$ or $H^+ \to \tau
\nu$~\cite{Tevtopdecays}.  In the lepton-specific 2HDM the partial
width for this top quark decay is proportional to $\cot^2\beta$, so
that the channel can be important only at low $\tan\beta \sim 1$; in
this parameter range the excluded regions can be taken over directly
from the usual Type-II 2HDM analysis.  The excluded regions lie below
$\tan\beta \simeq 2$ with $M_{H^+}$ between the LEP lower bound and
about 160~GeV~\cite{Tevtopdecays}.  This parameter region is already
excluded by the $b \to s \gamma$ constraint discussed in the previous
section.


\section{Charged Higgs branching fractions}
\label{sec:decaybrs}

We now present the decay branching fractions of $H^+$ in the
lepton-specific 2HDM, which we computed using a modified version of
the public FORTRAN code {\tt HDECAY}~\cite{hdecay}.  {\tt HDECAY}
computes the charged Higgs decay branching fractions in the Minimal
Supersymmetric Standard Model (MSSM), including decays to $\phi^0 W^{\pm}$
(with $\phi^0 = h^0$, $H^0$, or $A^0$) and supersymmetric particles when
kinematically accessible.  The Higgs sector of the MSSM has the Yukawa
coupling structure of a Type-II 2HDM.  

We adapt {\tt HDECAY} for the lepton-specific 2HDM by modifying the
charged Higgs couplings to fermions according to
Eq.~\ref{eq:Feynmanrules} and eliminating decays to supersymmetric
particles (no explicit supersymmetric radiative corrections to charged
Higgs decays are included in {\tt HDECAY}).  Decays to $\phi^0 W^{\pm}$ are
included; these decays depend on the scalar sector of the model and
their partial widths are the same in the lepton-specific 2HDM as in
the Type-II model for equivalent parameter sets.

\begin{figure}
\resizebox{0.45\textwidth}{!}{\includegraphics{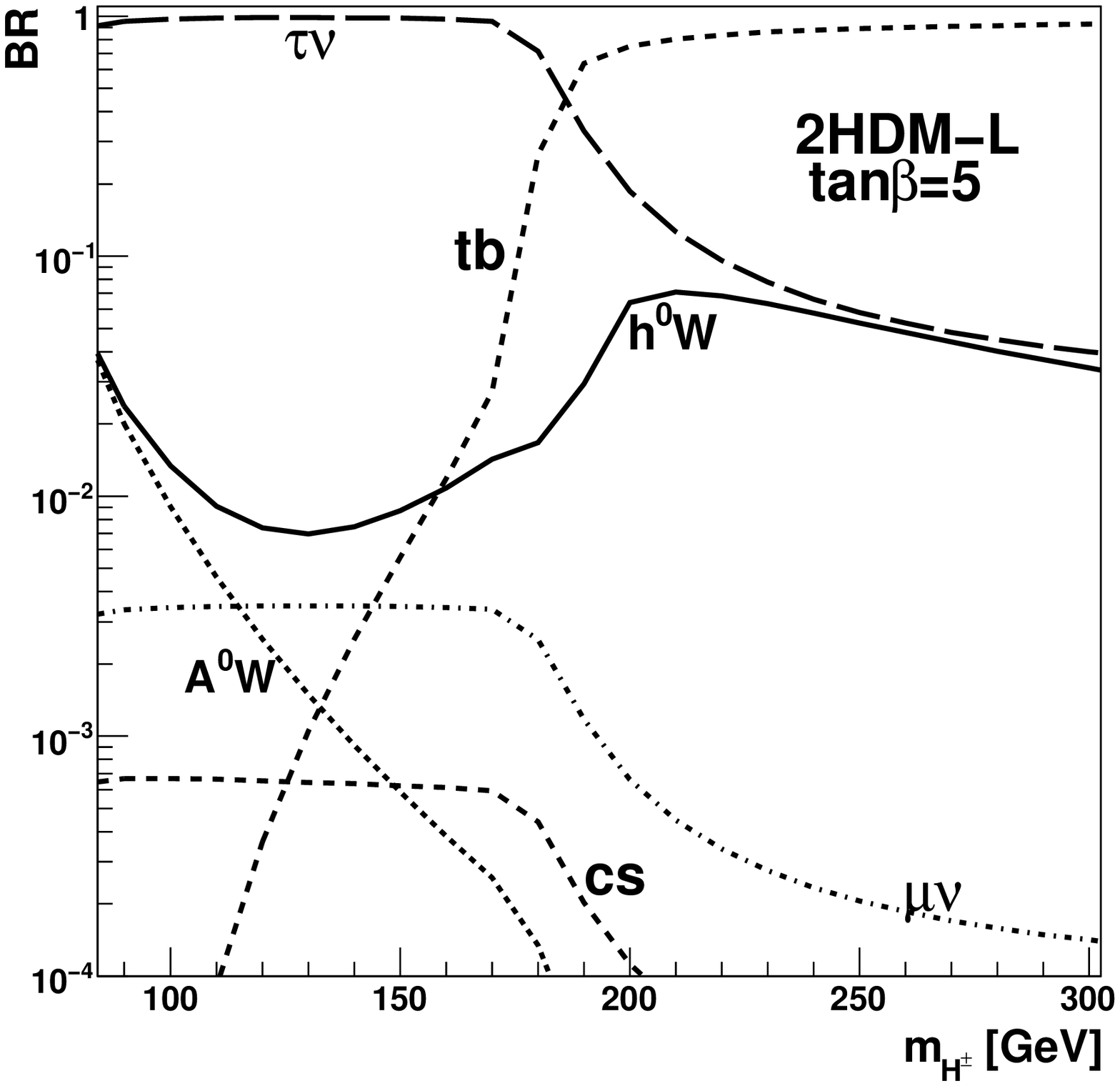}}
\resizebox{0.45\textwidth}{!}{\includegraphics{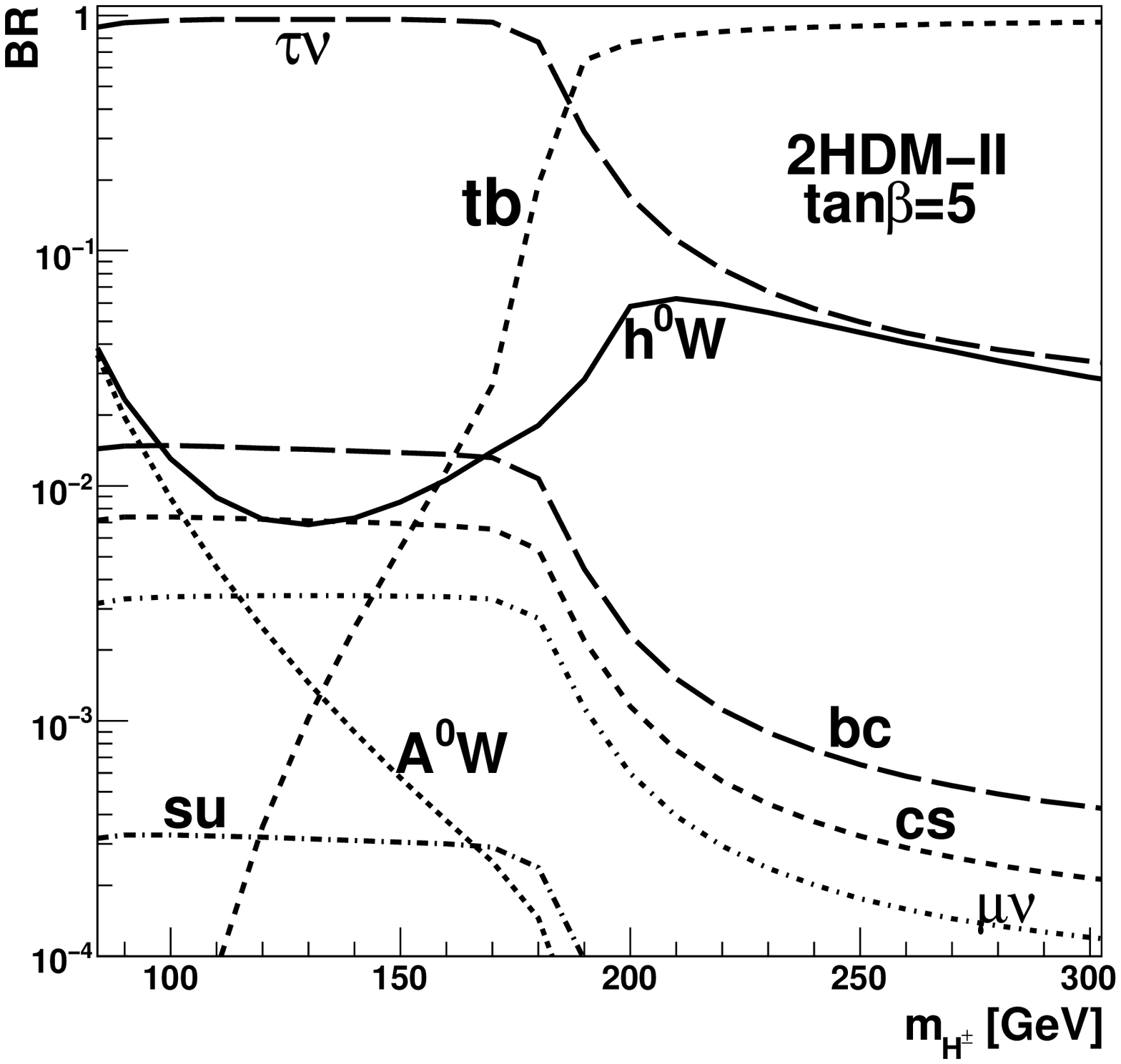}}
\caption{\label{fig:BR5}Branching ratios of $H^{\pm}$ as a function of
$M_{H^{\pm}}$ for $\tan\beta=5$.}
\end{figure}

\begin{figure}
\resizebox{0.45\textwidth}{!}{\includegraphics{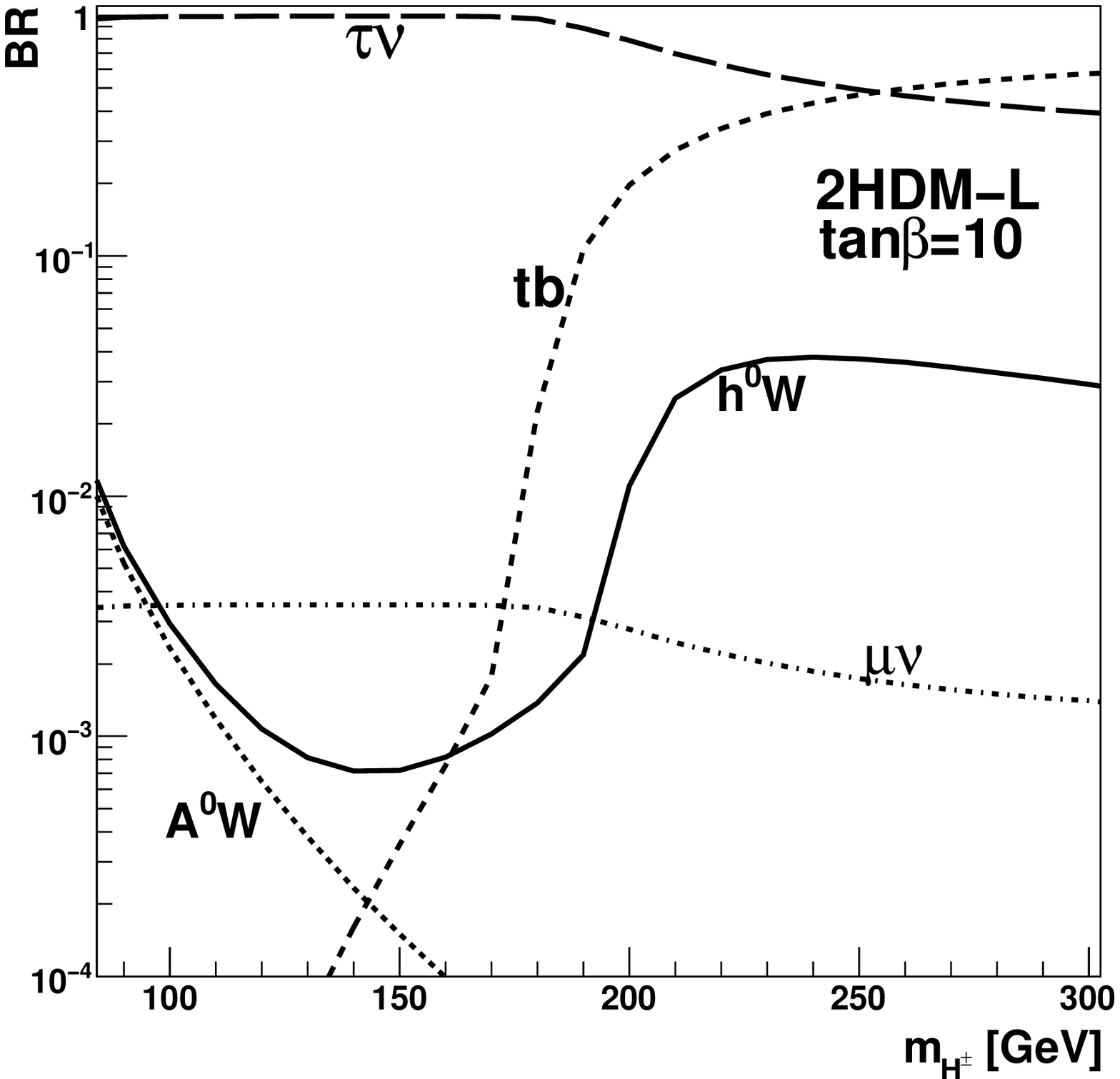}}
\resizebox{0.45\textwidth}{!}{\includegraphics{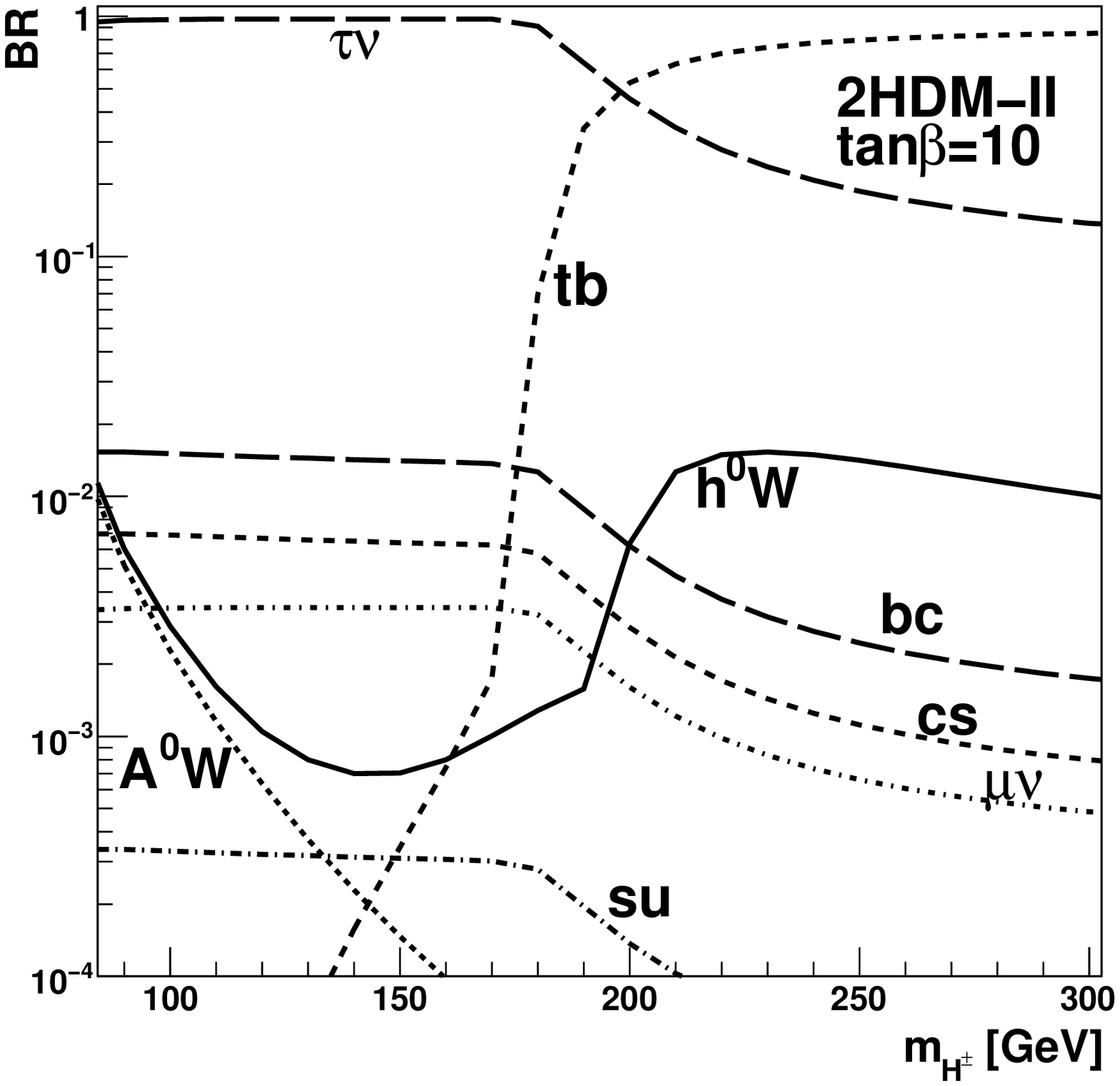}}
\caption{\label{fig:BR10}Branching ratios of $H^{\pm}$ as a function
of $M_{H^{\pm}}$ for $\tan\beta=10$.}
\end{figure}

\begin{figure}
\resizebox{0.45\textwidth}{!}{\includegraphics{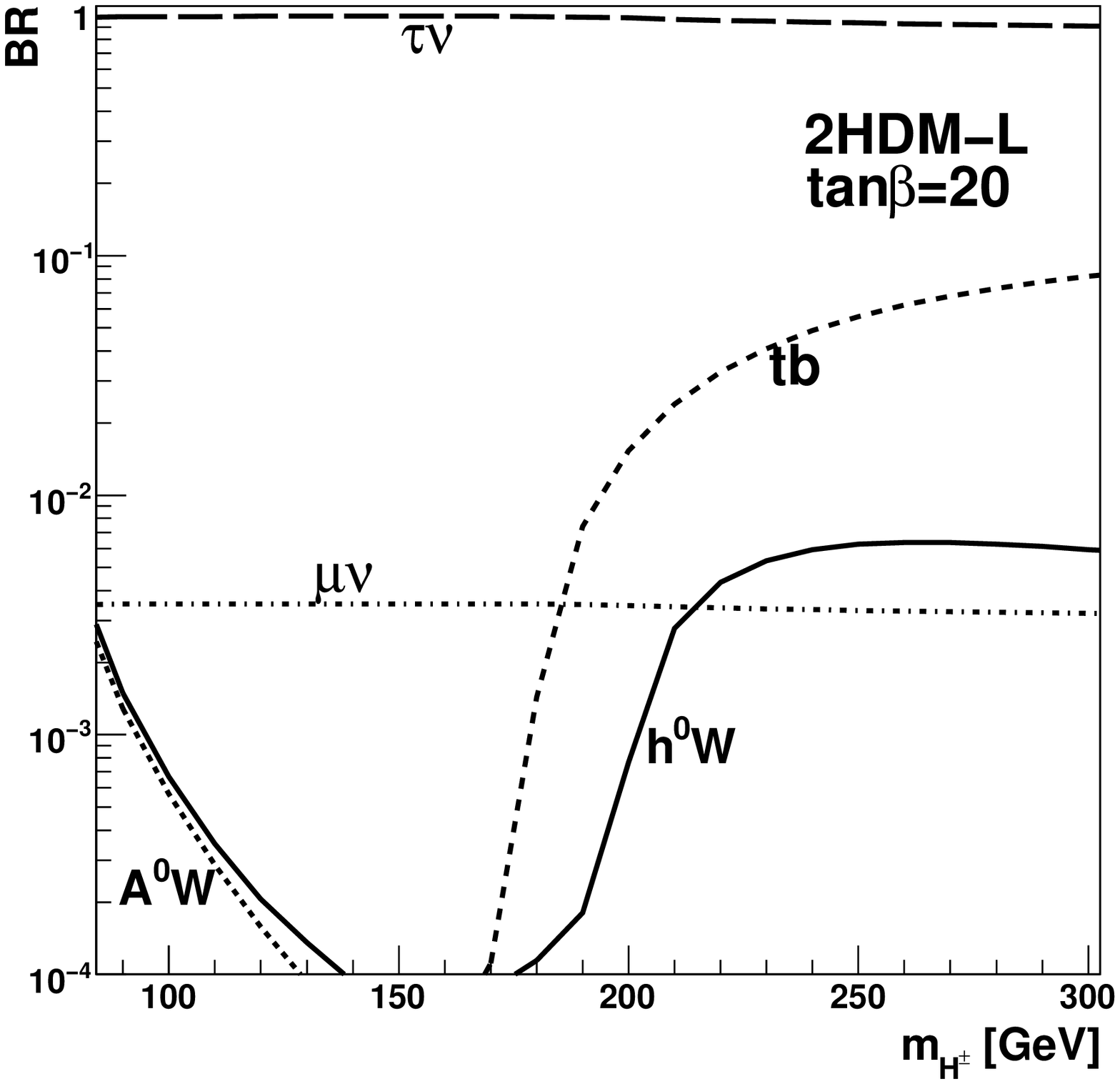}}
\resizebox{0.45\textwidth}{!}{\includegraphics{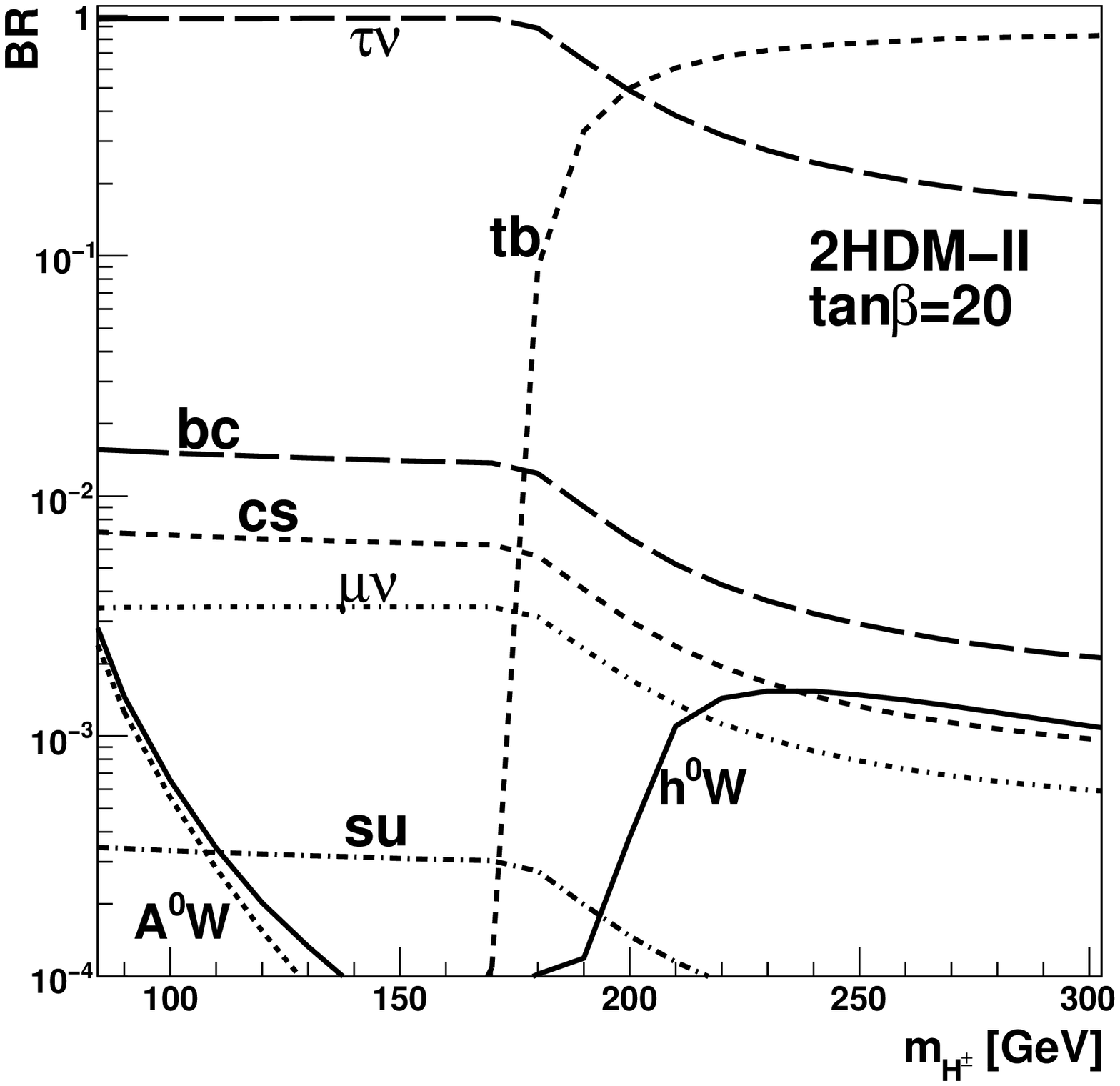}}
\caption{\label{fig:BR20}Branching ratios of $H^{\pm}$ as a function
of $M_{H^{\pm}}$ for $\tan\beta=20$.}
\end{figure}

\begin{figure}
\resizebox{0.45\textwidth}{!}{\includegraphics{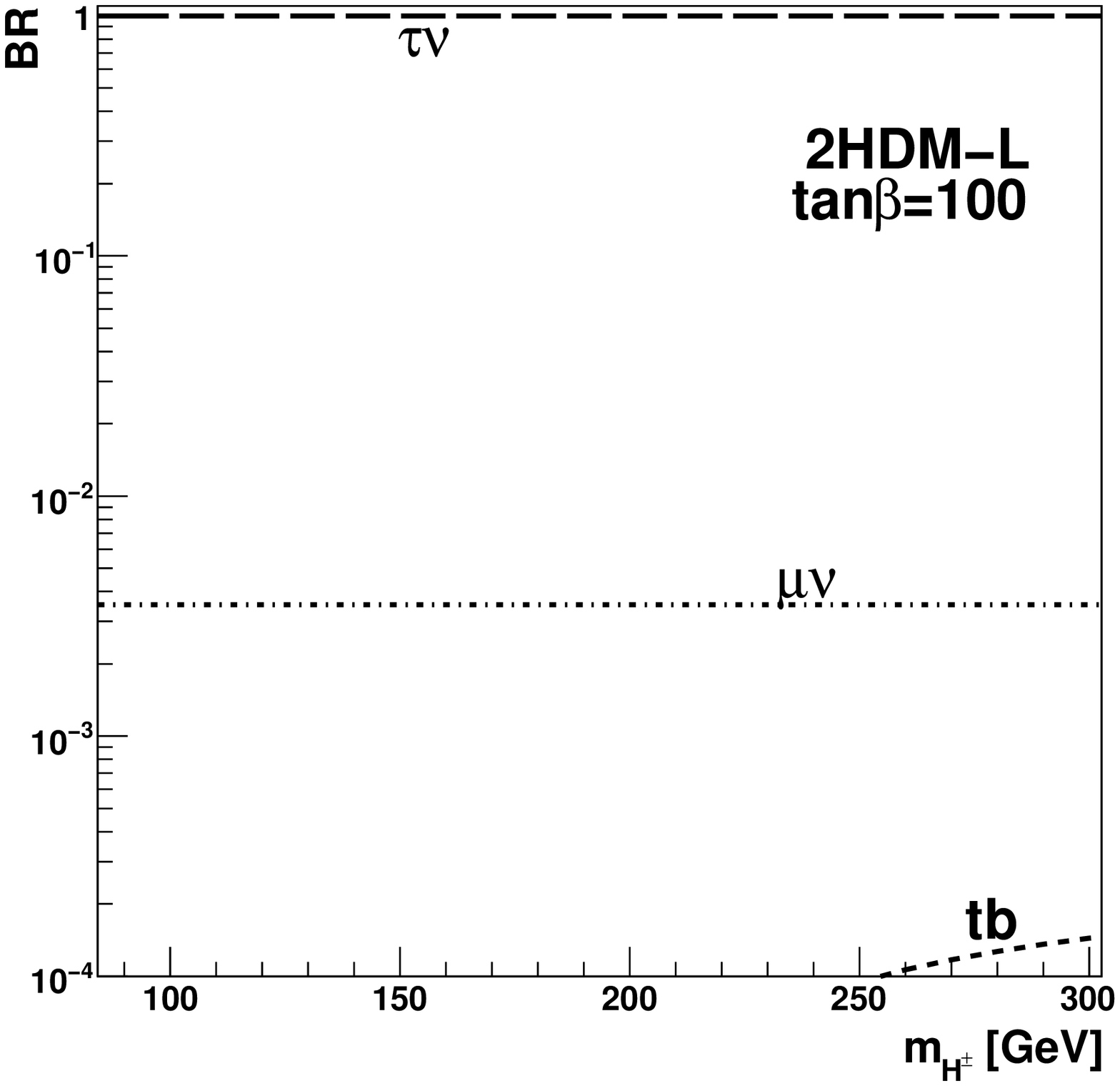}}
\resizebox{0.45\textwidth}{!}{\includegraphics{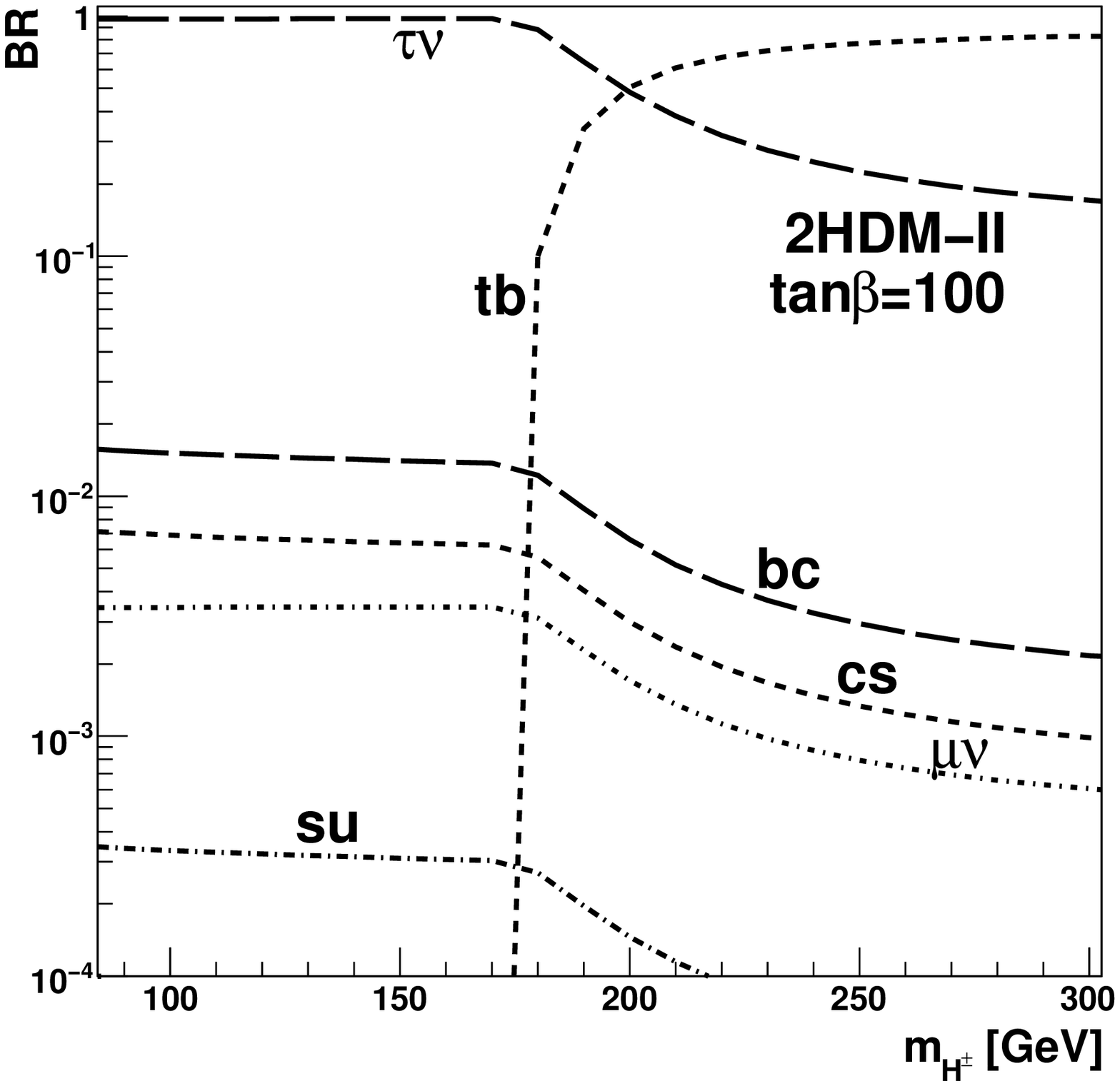}}
\caption{\label{fig:BR100}Branching ratios of $H^{\pm}$ as a function
of $M_{H^{\pm}}$ for $\tan\beta=100$.}
\end{figure}

In Figs.~\ref{fig:BR5}, \ref{fig:BR10}, \ref{fig:BR20} and
\ref{fig:BR100} we show the branching ratios of $H^{\pm}$ in the
lepton-specific 2HDM (2HDM-L) as a function of $M_{H^{\pm}}$ for
$\tan\beta = 5$, 10, 20, and 100, respectively.  For comparison we
also show the branching ratios of $H^{\pm}$ in the Type-II 2HDM
(2HDM-II).  For the decays to $A^0 W^{\pm}$ and $h^0 W^{\pm}$, we use
the $A^0$ and $h^0$ masses and the $h^0$--$H^0$ mixing angle predicted
in the MSSM as a function of $M_{H^{\pm}}$ and $\tan\beta$ with
all SUSY mass parameters set to 1~TeV.

For low $\tan\beta = 5$ (Fig.~\ref{fig:BR5}) the branching fractions
of $H^{\pm}$ in the lepton-specific 2HDM are quite similar to those in
the Type-II model, except that decays to $bc$ and $cs$ are suppressed.
This is due to the $\cot\beta$ suppression in the Yukawa couplings of 
both up- and down-type quarks in this model.  The $tb$ mode remains 
dominant for $M_{H^{\pm}} \gtrsim (m_t + m_b)$ because $m_t\cot\beta$ 
is still large compared to $m_{\tau}\tan\beta$ for $\tan\beta = 5$.

As $\tan\beta$ increases, the suppression of the quark modes becomes
more severe.  For $\tan\beta = 20$, the branching fraction to $\tau\nu$ 
remains above 90\% even for $M_{H^{\pm}}$ above the $tb$ threshold.
For higher $\tan\beta$ values, the leptonic decays dominate completely.

\begin{figure}
\resizebox{0.45\textwidth}{!}{\includegraphics{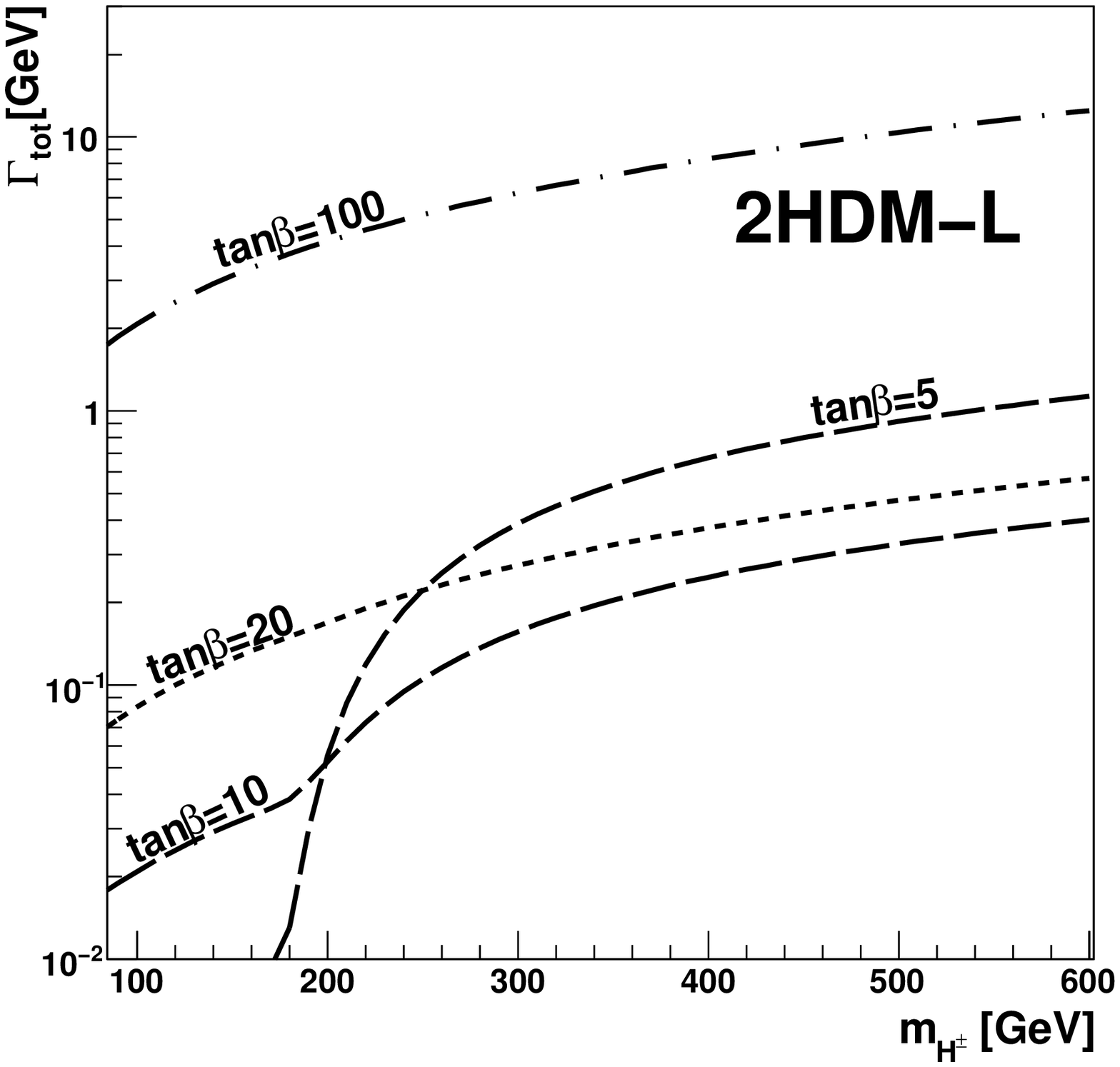}}
\resizebox{0.45\textwidth}{!}{\includegraphics{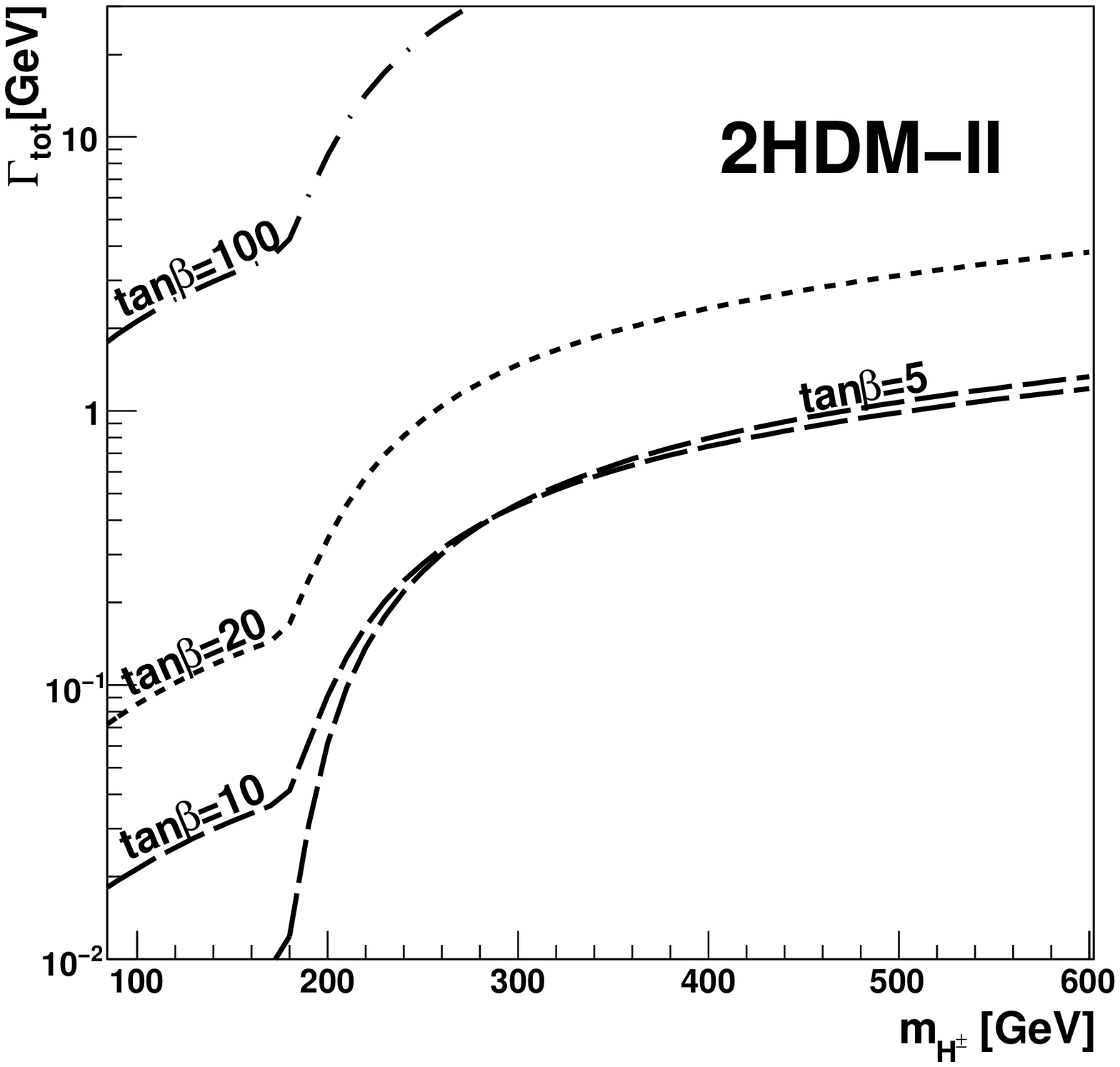}}
\caption{\label{fig:totalwidth}Total width of $H^{\pm}$ as a function
of $M_{H^{\pm}}$ for various values of $\tan\beta$.}
\end{figure}

In Fig.~\ref{fig:totalwidth} we show the total width of the charged
Higgs as a function of $M_{H^{\pm}}$, for $\tan\beta = 5$, 10, 20, and
100.  For comparison we again show the equivalent quantity for the
Type-II model.  Below the $tb$ threshold, where decays in both models
are dominated by the $\tau\nu$ final state, the total width of the
charged Higgs is comparable in the two models.  

Above the $tb$ threshold, however, the different Yukawa coupling
structure becomes obvious.  At low $\tan\beta = 5$ the total width is
dominated by $tb$ and the $tb$ threshold is obvious.  As $\tan\beta$
increases, however, the total width first declines, increasing again
only at large $\tan\beta$ where the $\tau\nu$ final state dominates
and the $tb$ threshold behaviour disappears entirely.  The total width
of the charged Higgs in the lepton-specific 2HDM remains quite
moderate, reaching $\sim 10$ GeV only for large $\tan\beta \sim 100$
at $M_{H^{\pm}} = 600$ GeV.  For lower $\tan\beta \sim 20$, the total
width remains below 1 GeV in this mass range, much lower than for the
Type-II model.


\section{Discussion and conclusions}
\label{sec:discussion}

The structure of the Yukawa couplings in the lepton-specific 2HDM
poses a challenge for charged Higgs discovery at the LHC.  The usual
LHC discovery channels for the charged Higgs of the MSSM or other
Type-II 2HDM involve production in association with a top
quark~\cite{tH-assocprod,tH-NLO} followed by decay to $\tau\nu$ or
$tb$~\cite{CMSTDR,ATLAS-CSCbook}.  In the MSSM this production channel
is particularly promising at large $\tan\beta$ because the production
cross section due to Yukawa radiation off the bottom quark grows with
$\tan^2\beta$.  In the lepton-specific 2HDM, however, the cross
section in this channel is proportional to $\cot^2\beta$ and thus
heavily suppressed at large $\tan\beta$.  
For $M_{H^{\pm}}$ below the top quark mass, the decay $t \to H^+ b$ with
$H^+ \to \tau\nu$ has also been studied for the
LHC~\cite{Baarmand:2006dm,ATLAS-CSCbook}.  In the lepton-specific 2HDM
the branching fraction for $t \to H^+ b$ is again suppressed by
$\cot^2\beta$.  We translate the $5\sigma$ charged Higgs discovery
sensitivity quoted in Ref.~\cite{ATLAS-CSCbook} into the
lepton-specific model by computing BR($t \to H^+ b$) at tree level; we
find the LHC discovery reach with 30~fb$^{-1}$ to be $\tan\beta
\lesssim 4.9$ (4.6, 2.4) for $M_{H^{\pm}} = 100$ (120, 150) GeV.
Likewise, all other
bottom-parton induced charged Higgs production processes in this
model, such as $H^+W^-$ associated production~\cite{H+W-} and $b\bar b
\to H^+ H^-$~\cite{BarrientosBendezu:1999gp,Alves:2005kr}, as well as
gluon fusion production of $H^+H^-$ via a third-generation quark
loop~\cite{BarrientosBendezu:1999gp,ggH+H-}, are suppressed by powers
of $\cot\beta$.  Because of this, LHC searches for the charged Higgs
in the lepton-specific 2HDM will have to rely on other production
processes.

\begin{figure}
\resizebox{0.8\textwidth}{!}{\rotatebox{270}{\includegraphics{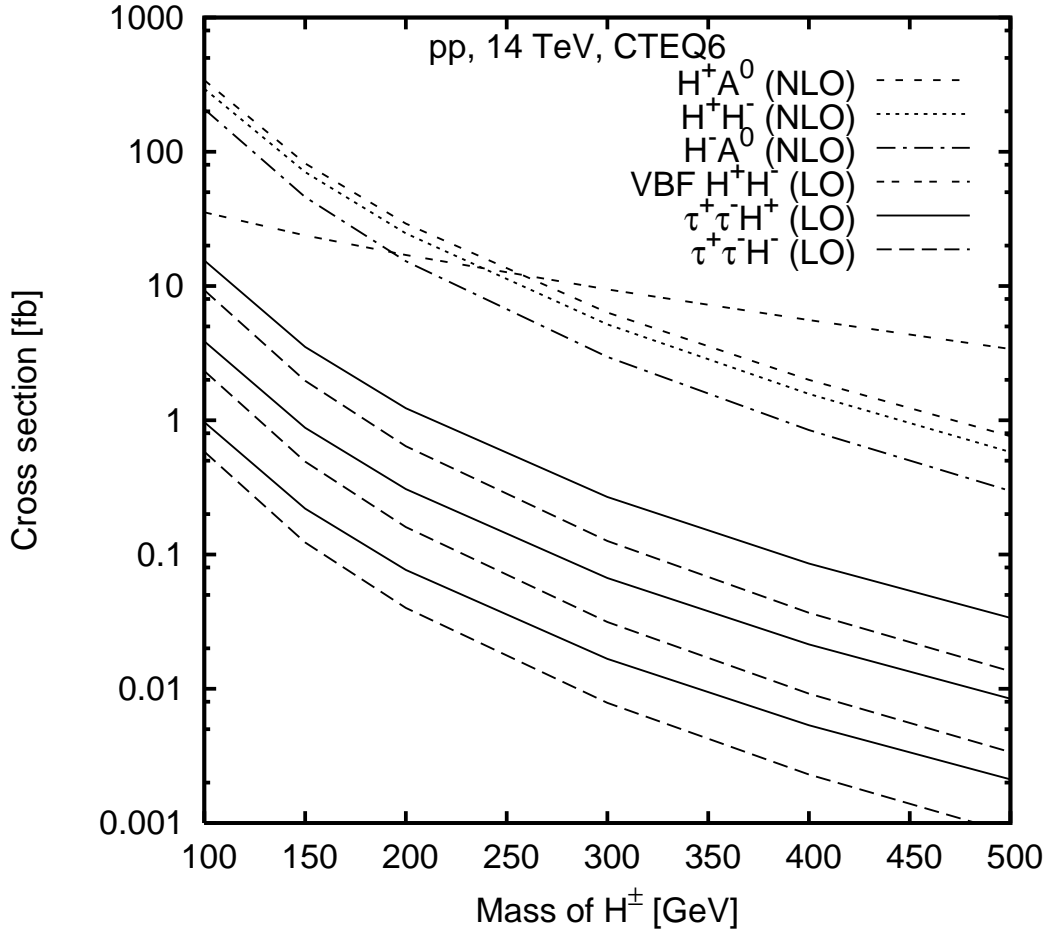}}}
\caption{\label{fig:xsec} Cross sections for charged Higgs production
at the LHC (see text for details).  The solid (dashed) lines show the
cross sections for $\tau^+ \tau^- H^+$ ($\tau^+ \tau^- H^-$)
production via Yukawa radiation for $\tan\beta = 200$, 100, and 50
from top to bottom.  For $H^{\pm} A^0$ associated production we take
$M_{A^0} = M_{H^{\pm}}$; the cross sections for $H^{\pm} H^0$ are
identical to those for $H^{\pm} A^0$ when $M_{H^0} = M_{A^0}$ and the
mixing angle in the CP-even sector is chosen so that the $W^+H^-h^0$
coupling vanishes.  LO (NLO) cross sections are computed using CTEQ6L
(CTEQ6M)~\cite{Pumplin:2002vw} with renormalization and factorization scales
set to $M_Z$ ($M_{H^{\pm}}$).}
\end{figure}

In Fig.~\ref{fig:xsec} we show the cross sections for various charged
Higgs production processes at the LHC.  Production of charged Higgs
pairs $q \bar q \to H^+H^-$ through an s-channel $Z$ or
photon~\cite{Eichten:1984eu,Alves:2005kr} depends only on the charged
Higgs mass once the SU(2) quantum numbers of the Higgs doublet are
fixed.  Similarly, associated production of $H^{\pm}$ and the CP-odd
neutral Higgs boson $A^0$ through an s-channel $W$
boson~\cite{Kanemura:2001hz,Cao:2003tr} depends only on the relevant
scalar masses.  Associated production of $H^{\pm}$ with a CP-even
neutral Higgs boson ($h^0$ or $H^0$) depends on the masses involved as
well as the mixing angle in the CP-even Higgs sector; if this mixing
angle is chosen such that the $W^+H^-h^0$ coupling vanishes, the
$H^{\pm} H^0$ cross section is equal to that for $H^{\pm} A^0$ for
degenerate $H^0$ and $A^0$.  We plot these cross sections in
Fig.~\ref{fig:xsec} including next-to-leading-order (NLO) QCD
corrections, computed using {\tt
PROSPINO}~\cite{Beenakker:1999xh}.\footnote{{\tt PROSPINO} computes
the cross sections for supersymmetric particle pair production at NLO.
We note that the $pp \to H^+H^-$ cross section is identical to that
for selectron pair production, $pp \to \widetilde e_L \widetilde
e_L^*$, and that the $pp \to H^+ A^0$ cross section is exactly half
that of $pp \to \widetilde e_L^* \widetilde \nu_e$ for corresponding
scalar masses.  We eliminate the supersymmetric QCD corrections
included in {\tt PROSPINO} by taking the squark masses to be very
heavy.}

Charged Higgs pair production due to vector boson fusion (VBF), $qq
\to qq V^*V^* \to qqH^+H^-$ ($V = \gamma$, $Z$, $W^{\pm}$), was
studied in detail in Ref.~\cite{Moretti:2001pp} in the MSSM.  The
cross section does not depend on the Yukawa structure of the model.
It is smaller than that for $q \bar q \to H^+ H^-$ for $M_{H^{\pm}}
\lesssim 250$ GeV; however, the two forward jets provide a powerful
selection tool against QCD backgrounds.  Ref.~\cite{Moretti:2001pp}
studied signal and backgrounds in the decay channel $H^+H^- \to
tb\tau\nu$ with the top quark decaying hadronically and found that the
QCD top-pair background remains overwhelming.  In the lepton-specific
2HDM, the dominant channel will be $H^+H^- \to \tau\nu\tau\nu$, which
may provide a cleaner signature.  We show this cross section in
Fig.~\ref{fig:xsec} as computed by {\tt
MadGraph/MadEvent}~\cite{Alwall:2007st}.\footnote{We impose the
following basic cuts on the jets in $pp \to jj H^+H^-$: $p_{Tj} \geq
20$~GeV, $\eta_j \leq 5$, $\Delta R_{jj} \geq 0.4$, and the dijet
invariant mass $m_{jj} \geq 100$~GeV.  Also, while their effects are
small~\cite{Moretti:2001pp}, neutral Higgs bosons enter as
intermediate states in the VBF $H^+H^-$ cross section calculation.
For the relevant masses we choose $M_{A^0} = M_{H^0} = M_{H^{\pm}}$
and $M_{h^0} = 120$ GeV.  We choose the mixing angle in the CP-even
sector so that the $W^+H^-h^0$ coupling vanishes.  The remaining free
parameter is the $h^0 H^+ H^-$ coupling; we choose the coefficient of
the Lagrangian term for $h^0H^+H^-$ to be equal to that for
$h^0h^0h^0$ for the given $h^0$ mass.}

Finally we consider the process $pp \to \tau^+ \tau^- H^{\pm}$ in
which the charged Higgs is radiated off one of the final-state $\tau$
leptons.  The squared matrix element for $\bar q q^{\prime} \to W^{+*}
\to \tau^+ \tau^- H^+$, neglecting external fermion masses, is given by
\begin{equation}
  \sum_{\rm spins} | \mathcal{M} |^2 = 
  g^4 \left[ \frac{g m_{\tau}}{\sqrt{2} M_W} \tan\beta \right]^2 
  \frac{4 \, p_2 \cdot k_1 \, 
    [2 k_2 \cdot k_3 \, p_1 \cdot k_3 - M_{H^{\pm}}^2 \, p_1 \cdot k_2]}
  {(q^2 - M_W^2)^2 \, (2 k_2 \cdot k_3 + M_{H^{\pm}}^2)^2},
\end{equation}
where $p_1$, $p_2$, $k_1$, $k_2$, and $k_3$ are the four-momenta of the 
incoming $\bar q$ and $q^{\prime}$, and outgoing $\tau^+$, $\tau^-$, and $H^+$,
respectively, and $q = p_1 + p_2$.
The cross section is proportional to $\tan^2\beta$; we show results
for $\tan\beta = 50$, 100, and 200 in Fig.~\ref{fig:xsec}, computed
using {\tt MadGraph/MadEvent}~\cite{Alwall:2007st}.

In summary, we studied the phenomenology of the charged Higgs boson in
the lepton-specific 2HDM.  We showed that the charged Higgs mass and
$\tan\beta$ are constrained by existing data from direct searches at
LEP and lepton flavour universality in $\tau$ decays; the former
yields $M_{H^{\pm}} \geq 92.0$~GeV and the latter yields two allowed
regions, $0.61 \tan\beta \ {\rm GeV} \leq M_{H^{\pm}} \leq 0.73
\tan\beta$ GeV or $M_{H^{\pm}} \geq 1.4 \tan\beta$~GeV, excluding
parameter space beyond the LEP-II bound for $\tan\beta \gtrsim 65$.
Improvements on $\tau$ decay branching fractions at the proposed
SuperB high-luminosity flavour factory would bring this reach down to
$\tan\beta \gtrsim 30$.  The $B$ meson decays that are usually used to
constrain the charged Higgs in the Type-II 2HDM provide no significant
constraints in the lepton-specific model because the charged Higgs
couplings to quarks are all proportional to $\cot\beta$.

We also studied the decay branching ratios of the charged Higgs in
this model and showed that decays to quarks are heavily suppressed at
large $\tan\beta$; in particular, the $t \bar b$ mode which typically
dominates above threshold in the Type-II 2HDM falls below the 10\%
level for $\tan\beta \gtrsim 20$.  Instead, $H^+ \to \tau \nu$
dominates at large $\tan\beta$ for all $H^+$ masses.  

The suppression of the quark couplings to the charged Higgs at large
$\tan\beta$ in this model poses a challenge for LHC discovery since it
suppresses the $tH^-$ associated production mode usually studied for
the Type-II 2HDM.  Instead, searches will have to rely on electroweak
production of $H^+H^-$ pairs or associated production of $H^{\pm}$
with a neutral Higgs boson.  The cross section for associated
production of $H^{\pm} \tau^+ \tau^-$ via Yukawa radiation is small
but it provides direct sensitivity to the $\tau$ Yukawa coupling.


\begin{acknowledgments}
This work was supported by the Natural Sciences and Engineering
Research Council of Canada.  We thank David Asner for helpful 
discussions on SuperB prospects.
\end{acknowledgments}

\vspace*{1cm} {\it Note added:} As this paper was being completed,
Ref.~\cite{Aoki:2009ha} appeared in which phenomenology of the same
model was studied.  Our results are largely consistent with theirs.
For the indirect constraint from $\tau$ decays we choose to use the
ratio of rates of $\tau \to \mu \nu \nu$ to $\tau \to e \nu \nu$ as
opposed to the partial width $\Gamma(\tau \to \mu\nu\nu)$ for two
reasons: (i) The experimental uncertainty on the ratio is smaller than
that on the partial width, due to the non-negligible uncertainty in
the $\tau$ lifetime; and (ii) the partial width $\Gamma(\tau \to
\mu\nu\nu)$ receives potentially significant one-loop contributions
from diagrams involving neutral Higgs bosons as pointed out in
Ref.~\cite{Krawczyk:2004na}; these effects cancel in the ratio of
rates, allowing direct sensitivity to the charged Higgs sector.



\end{document}